\pdfoutput=1
\documentclass{article}

\usepackage{preamble}

\newcolumntype{d}[1]{D{.}{.}{#1}}
\usepackage[nonatbib,final]{neurips_2022_ml4ps}

\title{PELICAN: Permutation Equivariant and Lorentz Invariant or Covariant Aggregator Network for Particle Physics}

\author{%
  Alexander Bogatskiy\\
  Center for Computational Mathematics\\
  Flatiron Institute, New York, NY, U.S.A.\\
  \texttt{abogatskiy@flatironinstitute.org}\\
  \And
  Timothy Hoffman\\
  Department of Physics, University of Chicago\\
  Chicago, IL, U.S.A.\\
  \texttt{hoffmant@uchicago.edu}\\
  \And
  David W.~Miller\\
  Department of Physics, University of Chicago\\
  Enrico Fermi Institute\\
  Chicago, IL, U.S.A.\\
  \texttt{David.W.Miller@uchicago.edu}\\
  \And
  Jan T.~Offermann\\
  Department of Physics, University of Chicago\\
  Enrico Fermi Institute\\
  Chicago, IL, U.S.A.\\
  \texttt{jano@uchicago.edu}\\
}

\begin{document}

\maketitle

\begin{abstract}
  Many current approaches to machine learning in particle physics use generic architectures that require large numbers of parameters, often adapted from unrelated data science or industry applications, and disregard underlying physics principles, thereby limiting their applicability as scientific modeling tools. In this work, we present a machine learning architecture that uses a set of inputs maximally reduced with respect to the full 6-dimensional Lorentz symmetry, and is fully permutation-equivariant throughout. We study the application of this network architecture to the standard task of classifying the origin of jets produced by either hadronically-decaying massive top quarks or light quarks, and show that the resulting network outperforms all existing competitors despite significantly lower model complexity. In addition, we present a Lorentz-covariant variant of the same network applied to a 4-momentum regression task in which we predict the full 4-vector of the W boson from a top quark decay process.
\end{abstract}

\section{Introduction}
\label{Introduction}

    Neural networks have played a significant role in data analysis for particle physics experiments, perhaps most significantly in the study of \textit{jets}, the collimated streams of hadronic particles produced by the decay, showering and hadronization of quarks and gluons. As demonstrated in Ref.~\cite{Cogan:2014oua}, convolutional neural networks can be leveraged for identifying the type of particle that initiated a jet. However, such network architectures do not explicitly respect (or leverage) the symmetries inherent in particle physics, in particular those of the \textit{Lorentz group}. Designing a network architecture that respects these symmetries may benefit model interpretability while reducing model complexity via physically-meaningful constraints, without sacrificing performance.

\section{Equivariance and jet physics}
\label{equivariance}

\paragraph{Lorentz Invariance}\label{LI}

In what follows, ``Lorentz group'' refers to the \textit{proper orthochronous Lorentz group} $\SO^+(1,3)$, i.e.~the identity component of the full Lorentz group $\mathrm{O}(1,3)$ of linear transformations on $\bbR^4$ that preserve the Minkowski metric $\eta=\mathrm{diag}(1,-1,-1,-1)$. The classification task considered in this work is Lorentz invariant, that is, the output of the network is invariant under the application of any Lorentz transformation $\Lambda\in\SO^+(1,3)$ to all of the 4-vector inputs (energy-momentum vectors in our case, for which  $p=(p^0,\vec{p})=(E,p^x,p^y,p^z)$ and $p^2=E^2-\vec{p}^2$).
The simplest way to enforce invariance is to hand-pick a set of invariant observables (e.g. particle masses, identification labels) as inputs to a generic neural network architecture, as summarized in Ref.~\cite{KasiePlehn19}.  Another approach inspired by convolutional neural networks (CNN's) is to preserve group-equivariant latent representations in the hidden layers, see e.g.~Refs.~\cite{Bogatskiy:2020tje, LorentzNet22}. As in CNN's, equivariant latent representations, as opposed to invariant ones, can regularize the network via efficient weight-sharing ~\cite{ExponentialSep}.

Here, we take a slightly different approach. Given a set of 4-vector inputs $p_1,\ldots,p_N$, we compute the \textit{complete} set of Lorentz invariants on that set. Weyl's work \cite{Weyl46} characterized the set of all Lorentz invariant functions of a collection of 4-vector inputs. Namely, all totally symmetric Lorentz invariants $I(p_1,\ldots,p_N)$ depend only on the invariant dot products (see related discussion in Ref.~\cite{Hogg}):
\[I(p_1,\ldots,p_N)=I\left(\{p_i\cdot p_j\}_{i,j}\right).\label{eq}\]
The array of all $N\times N$ pairwise dot products will be the network input. We note that the idea to include dot products as inputs was recently used in Ref.~\cite{DoesLorentzProvide}. As we will show, these invariant dot products alone can provide state-of-the-art performance in a significantly simpler architecture.

\paragraph{Permutation Equivariance}\label{EUA}

Particle data is often naturally represented by a point cloud, or a set. For such problems the ordering of the particles in the set is not physically meaningful, and thus it makes sense to use one of the permutation-equivariant architectures. One approach is that of Deep Sets \cite{ZaKoRaPoSS17}, applied to jet tagging e.g.~in Ref.~\cite{EFN}. It is based on the fact that any symmetric function of inputs $x_1,\ldots,x_N$ can be written in the form $\psi\left(\sum_i\varphi(x_i)\right)$, where $\psi$ and $\varphi$ can be approximated by neural networks. However, since aggregation happens only once, the network can struggle at modeling complex higher-order interactions between the particles. The sub-network representing $\psi$ is forced to be a relatively complex (wide) fully-connected network, which makes it difficult to train \cite{Wagstaff19, ExponentialSep}. The alternative to permutation-invariant architectures is provided by permutation-\textit{equivariant} ones. 
Equivariance is a key property of all convolutional networks -- for example, in CNN's convolutions are manifestly equivariant with respect to translations (up to edge effects). Similarly, Graph Neural Networks (GNN's) use permutation equivariance, usually in the form of message passing (MP), to force architectures to respect the underlying graph structure. 

Despite the benefits of MP (previously used in jet tagging \cite{Bogatskiy:2020tje, LorentzNet22}), attempts to combine MP with Lorentz invariance run into an obstacle: the key inputs to the network are \textit{nothing but} edge data $d_{ij}=p_i\cdot p_j$. Since traditional MP architectures use only single-label vertices, we employ the general permutation-equivariant layers proposed in Refs.~\cite{Aggregators, KondorPan}. In the general setting, permutation equivariance is a constraint on mappings $F$ between arrays $T_{i_1i_2\cdots i_r}$ of any rank $r$, where every index $i_k\in \{1,\ldots,N\}$ refers to a particle label, whereby permutations $\pi\in S_N$ of the particles ``commute'' with the map:
\[F\left(\pi\circ T_{i_1i_2\cdots i_r}\right) =\pi\circ F\left(T_{i_1i_2\cdots i_s}\right), \quad \pi\in S_N.\]
Here, the action of permutations is diagonal: $\pi\circ T_{i_1i_2\cdots i_p} =T_{\pi(i_1)\ldots \pi(i_p)}$. Thus a higher-order generalization of the MP layer can be defined as $T^{(\ell+1)}=\textsc{Agg}\circ \textsc{Msg}\left(T^{(\ell)}\right)$. Here, $\textsc{Msg}$ is a node-wise nonlinear map (``message forming'') shared between all nodes, and $\textsc{Agg}$ is a general permutation-equivariant linear mapping (``aggregation'') acting on the particle indices of $T$. 

\paragraph{Elementary Equivariant Aggregators}\label{EEA}

It remains to describe the exact structure of the equivariant aggregation layers introduced above. Since the general case is presented in \cite{Aggregators, KondorPan}, here we will only present the layers needed for jet physics. Since the input is an array of rank $2$ (of dot products), the main equivariant layer in this case is one that maps between arrays of rank $2$: $T_{ij}\mapsto T_{ij}'$. The space of all linear maps of this type is 15-dimensional and its basis elements can be defined using binary arrays of rank $4$. There are 15 such arrays $B^{a}_{ijkl}, a=1,\ldots,15$, (see Ref.~\cite{KondorPan} for exact expressions) and the action of the equivariant layer can be written as $T^{\prime a}_{ij}=\sum_{k,l=1}^N B^{a}_{ijkl} T_{kl}$. Five of the basis elements in fact contain only one non-zero component for each $ij$ pair, which includes the identity and the transposition maps, so they can be thought of as sorts of ``skip connections''. The rest involve aggregations over $N$ or $N^2$ elements of the input.

More generally, instead of a simple summation, aggregators can involve arbitrary (nonlinear) symmetric functions, e.g.~maximum. In practice, we define aggregation as the mean of its inputs followed by an additional scaling by a factor of $\left(N/\bar{N}\right)^\alpha$ with learnable exponents $\alpha$, where $\bar{N}$ is a constant describing the typical number of particles expected in the input.

\paragraph{Equivariance and Jet Physics}

There are several reasons for enforcing the full Lorentz symmetry in our ML models. First and foremost, it is a fundamental symmetry of the space to which the inputs belong. 
If an analyzer in the lab frame establishes that a given collection of particles resulted from a top quark decay, then the same is true for all other reference frames. The breaking of the Lorentz symmetry implicit in the running of the QCD couplings notwithstanding, there is no question that both the original protons and the final (asymptotic) decay products are accurately represented by a collection of 4-vectors subject to the global Lorentz symmetry. Another reason for symmetry-restricted modeling is that, from the geometric perspective, only some mathematical operations are permissible when working with objects that transform in a certain way under a symmetry group. A non-equivariant neural network effectively neglects the vector nature of the inputs by treating individual components of the input vectors as scalars. Despite \textit{a priori} improving network expressivity, non-equivariance fails to deliver physically interpretable models. Ultimately, a statement about equivariance is a statement about what quantities are not truly self-contained \textit{features} of the inputs -- e.g.~a single $x$-component of a 2D vector is not a feature of that vector unless the input also contains the vector $(1,0)$.


\section{PELICAN architecture}
\label{architecture}

\paragraph{Permutation Equivariant Blocks}

\begin{wrapfigure}{r}{0.25\linewidth}
    \vspace{-0.8\intextsep}
    \centering
    \includegraphics[scale=0.7]{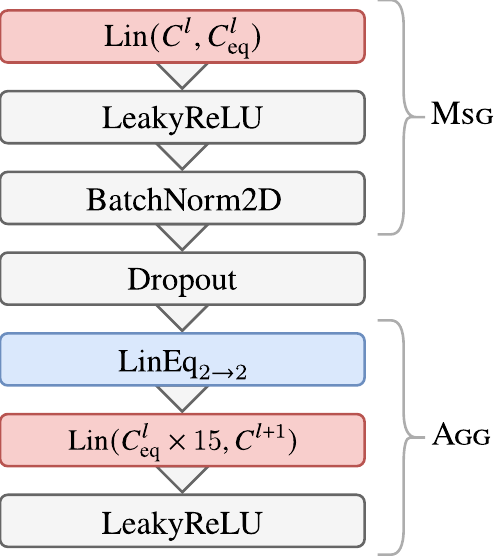}
    \vspace{-20pt}
\end{wrapfigure}


The main equivariant block, $\texttt{Eq}_{2\to 2}$, consists of a simple dense layer $\textsc{Msg}$ and an aggregation block $\textsc{Agg}$. The aggregation block applies 15 linear aggregation functions ($\texttt{LinEq}_{2\to 2}$) as outlined in Section \ref{EEA}. Note that this is a non-parametric transformation performed on each channel separately. Each of the $C_{\text{eq}}^l\times 15$  resulting aggregation values is then independently multiplied by $N^\alpha/\bar{N}^\alpha$ with a trainable exponent $\alpha$ (initialized as a random float in $[0,1]$, allowed to become negative), where $N$ is the number of particles in the corresponding event. This allows for some flexibility in the aggregation process, for example $\alpha=1$ returns the sum aggregation function, and combining multiple aggregators is known to boost accuracy \cite{Aggregators}. PELICAN stacks five such blocks for optimal results according to a loss-minimizing hyperparameter search.

\paragraph{PELICAN Classifier}
To build a classifier, aside from the $\texttt{Eq}_{2\to 2}$ equivariant layer one needs a $\texttt{Eq}_{2\to 0}$ layer that reduces the rank 2 array to permutation-invariant scalars. This layer involves just 2 aggregation functions instead of $15$ -- the trace and the total sum of the input square matrix, but is otherwise identical to the equivariant block described above. 
The inputs $d_{ij}$ are positive with a very steeply decaying distribution at large values, therefore after forming the matrix of pairwise dot products the input layer applies a set of encoding functions of the form $((1+x)^\delta-1)/\delta$, with $\delta=\beta^2$ and learnable $\beta$'s. From the input block, the tensor is passed through several equivariant $\texttt{Eq}_{2\to 2}$ blocks, and a $\texttt{Eq}_{2\to 0}$ block, all with dropout. One final dense layer mixes the channels down to $2$ classification weights per event. A cross-entropy loss function is then used for optimization.

\paragraph{PELICAN Regressor} The same architecture can also be easily adapted for 4-momentum regression tasks, such as momentum reconstruction. Any Lorentz-equivariant map from a collection of 4-momenta $p_1,\ldots,p_N$ to one (or several) 4-momentum has the form
\[F(p_1,\ldots,p_N)=\sum_{i=1}^N f_i(p_1,\ldots,p_N)\cdot p_i,\]
where $f_i$'s are Lorentz-invariant functions \cite{Hogg}. Combining this with permutation-invariance, we conclude that the multi-valued map $(p_1,\ldots,p_N)\mapsto (f_1,\ldots,f_N)$ must also be equivariant with respect to the permutations of the inputs. The only change required to the architecture we've introduced for classification is that  $\texttt{Eq}_{2\to 0}$ must be replaced with $\texttt{Eq}_{2\to 1}$ and the final output layer must have only one output channel. The $\texttt{Eq}_{2\to 1}$ layer is again identical to $\texttt{Eq}_{2\to 2}$ except that it uses only $4$ linear aggregation functions. For the loss function we use $5\left|m_p-m_t\right|+\left\vert \vec{p}_p-\vec{p}_t\right\vert$ where subscripts $p,t$ stand for predicted and true values, respectively. We avoid squares due to their sensitivity to outliers, and the coefficients of the two terms are chosen to roughly balance their magnitudes on our dataset, forcing the network to simultaneously predict the mass and the spatial momentum.

\section{Experiments}
\label{toptagging}

\paragraph{Top tagging}

We perform top-tagging on the reference dataset \cite{KasPleThRu19} explored in Ref.~\cite{KasiePlehn19}. This dataset consists of 2M entries, each entry corresponding to a single hadronic top jet or the leading jet from a QCD dijet event. The events were generated with \Pythia8.2~\cite{Sjostrand:2014zea}, with \Delphes~\cite{deFavereau:2013fsa} used for detector interactions. For each jet, the four-momenta of up to $200$ constituents are listed. The model was trained in batches of 100 events on A100 80GB GPU's using the AdamW optimizer \cite{AdamW} with a linear warmup of the learning rate up to $2.5\cdot 10^{-3}$ for the first 4 epochs, followed by 28 epochs of \texttt{CosineAnnealingWarmRestarts} with $T_0=4$, $T_\mathrm{mult}=2$, and 3 more epochs of exponential schedule with $\gamma=0.5$. A dropout rate of 1\% was used. The \textsc{Msg} blocks output 35 channels and the \textsc{Agg} blocks output 60. Hyperparameters were tuned by manual optimization.

\begin{wrapfigure}{r}{0.44\linewidth}
    \vspace{-1.4\intextsep}
    \centering
    \includegraphics[trim=10 10 10 10, clip, scale=0.33]{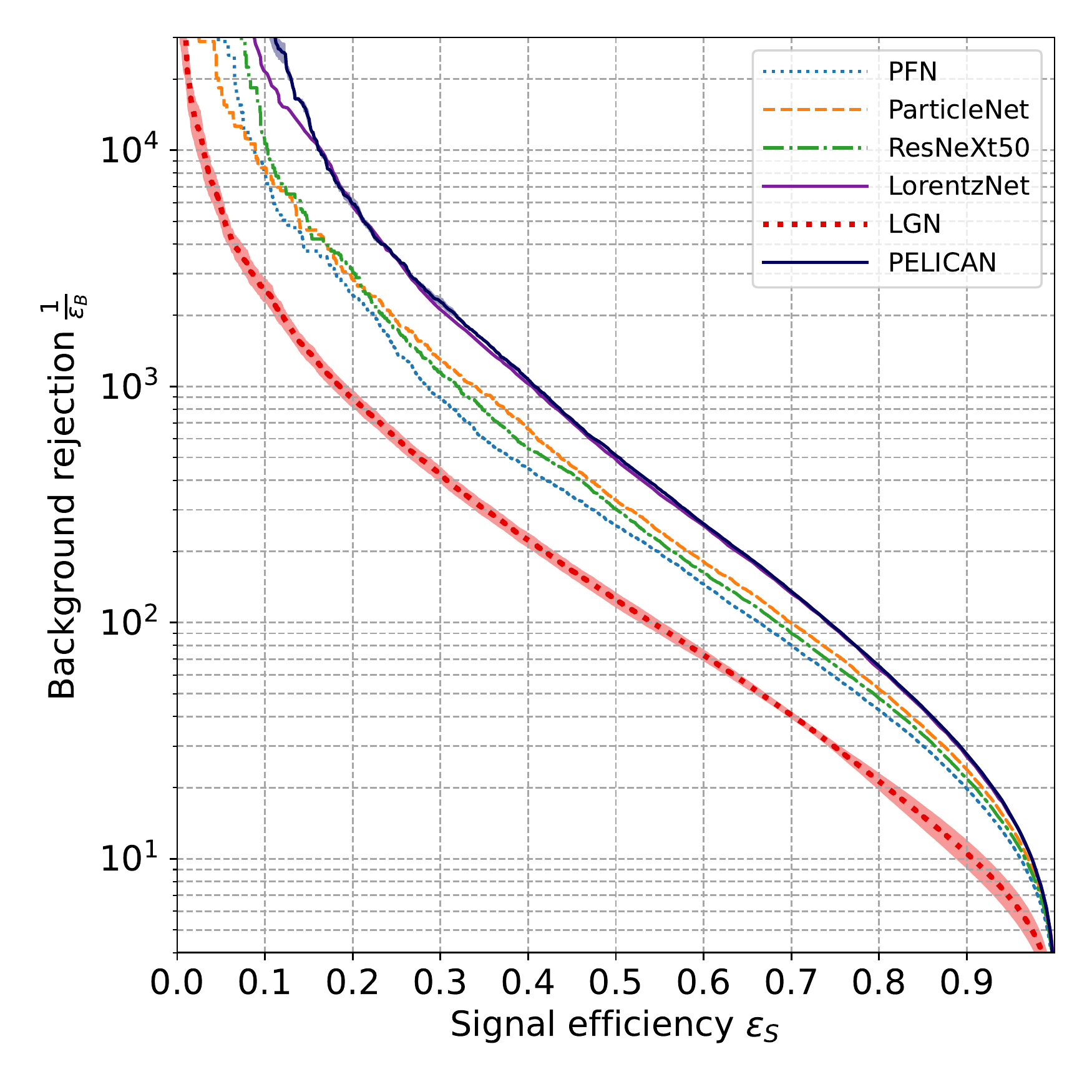}
    \caption{Top-tagger ROC curves.}
    \vspace{-20pt}
\end{wrapfigure}
The metrics of several top taggers are compared in Table~\ref{tab1}, quoted from Refs.~\cite{KasiePlehn19,LorentzNet22}. $1/\epsilon_B$ stands for the background rejection rate at efficiency rate of $0.3$. Among these PFN, LGN and LorentzNet are physics-motivated architectures, with LorentzNet using a Lorentz-invariant version of Message Passing. 
 
\begin{wraptable}{l}{0.7\linewidth} 
    \vspace{-0.6\intextsep}
    \caption{Comparison of top-taggers.}
    \label{tab1}
    \centering
    \begin{small}
    \begin{tabular}{l@{\hspace{2mm}}l@{\hspace{2mm}}l@{\hspace{2mm}}l@{\hspace{2mm}}r}
    \toprule
    Architecture    &   Accuracy    &   AUC         &   $1/\epsilon_B$  &   \# Params \\
    \midrule
    LGN             &   0.929(1)    &   0.964(14)   & 424 $\pm$ 82      &   4.5k    \\
    PFN             &   0.932       &   0.982       &   891 $\pm$ 18    &   82k     \\
    ResNeXt         &   0.936       &   0.984       &   1122 $\pm$ 47   &   1.46M   \\
    ParticleNet     &   0.938       &   0.985       &   1298 $\pm$ 46   &   498k    \\
    LorentzNet      &   0.942       &   0.9868      & 2195 $\pm$ 173    &   220k    \\     
    \midrule
    PELICAN         &   0.9425(1)   &   0.9869(1)   & 2289 $\pm$ 204    &   45k     \\   
    \bottomrule
    \end{tabular}
    \end{small}
\end{wraptable}
\WFclear 

Our results are averaged over 5 random initialization seeds and the uncertainties are given by the standard deviation. As can be seen from these results, equivariant architectures provide good models for top-tagging despite much lower model complexity as measured by the number of trainable parameters. PELICAN in particular provides state-of-the-art accuracy with almost 5 times fewer parameters than the next best model. Model complexity is of great importance for many real-world applications, such as online detector triggering, which requires extremely low microsecond latencies \cite{OneShot, BDTinHEP, Triggering, GNNinReconstruction, LowLatency}. Normally reduction in model size comes at the cost of accuracy, but equivariant architectures can avoid such compromises. Aside from accuracy, physics applications also demand high background rejection rates, where PELICAN also provides state-of-the-art performance. 

\renewcommand{\arraystretch}{1.05} 
\begin{wraptable}{r}{0.6\linewidth}
    \vspace{-1\intextsep}
    \caption{Momentum reconstruction results for the Johns Hopkins (JH) tagger and PELICAN. We report the relative $p_T$ and mass resolutions, and the interquantile range for the angle $\psi\in(0,\pi)$ between predicted and true momenta. PELICAN uncertainties are within the last significant digit.}
    \label{tab2}
    \centering
    \begin{small}
        \begin{tabular}{ccS[table-format=3.2]<{\%}S[table-format=3.2]<{\%}S[table-format=3.3]}
        \toprule
        & Method &  \multicolumn{1}{c}{$\sigma_{p_T}$ (\%)} & \multicolumn{1}{c}{$\sigma_{m}$ (\%)} & \multicolumn{1}{c}{$\sigma_\psi$ (centirad)}\\
        \midrule
        \multirow{3}{*}{\rotatebox[origin=c]{90}{\parbox{1.3cm}{\centering Without\\ \Delphes}}} 
        & JH               & 0.70    & 1.29     & 0.162   \\
        & PELICAN          & 0.83    & 1.21     & 0.388   \\
        & PELICAN$\mid$JH  & 0.28    & 0.60     & 0.089   \\
        & PELICAN$\mid$FC  & 0.32    & 0.76     & 0.111   \\
        \midrule
        \multirow{3}{*}{\rotatebox[origin=c]{90}{\parbox{1.3cm}{\centering With\\ \Delphes}}} 
        & JH               & 10.8   & 8.3   & 8.9      \\
        & PELICAN          & 5.6    & 3.2   & 4.2      \\
        & PELICAN$\mid$JH  & 3.8    & 2.9   & 2.7      \\
        & PELICAN$\mid$FC  & 4.4    & 3.1   & 3.0      \\
        \bottomrule
        \end{tabular}
    \end{small}
    \vspace{-0.5\intextsep}
\end{wraptable}

\paragraph{4-Momentum reconstruction}
For regression, we use a dataset~\cite{offermann_data_2022} containing 1M entries, each corresponding to a single hadronic top jet from an event generated with \Pythia8, with settings as in Ref.~\cite{KasPleThRu19}. We prepare two versions of this dataset: One with jets clustered from truth-level, final-state particles, the other clustered from the output of detector simulation with \Delphes, each using the same events. Each entry contains the $200$ leading jet constituents, and the parton-level 4-momenta of the top quark, and of the b-quark and $W$-boson to which it decays.

Our regression task consists of predicting the 4-momentum of the parton-level $W$-boson in the lab frame. The results are summarized in Table \ref{tab2} by the resulting $p_T$ and mass resolutions -- given by half of the central $68^\text{th}$ interquantile range of $(x_{\mathrm{predict}}-x_{\mathrm{true}})/x_{\mathrm{true}}$, where $x$ is $m$ or $p_T$ -- and the lower $68^\text{th}$ interquantile range for $\psi$, the angle between predicted and true momenta. To serve as a baseline regression method, we use the $W$-boson identification of the Johns Hopkins top tagger~\cite{Kaplan:2008ie} implemented in \Fastjet~\cite{Cacciari:2011ma}. The tagger has a $36\%$ efficiency on the dataset and can only identify $W$-boson candidates for jets it tags, so we report PELICAN results both on the tagged jets (PELICAN$\mid$JH) and on the full dataset. In addition, we report PELICAN results on the subset of jets that are \textit{fully-contained} (PELICAN$\mid$FC), which we define as jets where both quarks from $W$-boson decay are within the jet radius. This is in fact a strict subset of the tagged jets, and we highlight this subpopulation of jets as we do not expect accurate $W$-boson momentum reconstruction in the case of jets that fail to capture a significant fraction of the $W$-boson decay products. Notably, fully-contained events comprise about 75\% of the dataset, which is still significantly higher than JH tagger's efficiency. The regression network uses 36k parameters and was trained in the same way as the classifier. The code for PELICAN can be found at \href{https://github.com/abogatskiy/PELICAN}{github.com/abogatskiy/PELICAN}.

\paragraph{Conclusion}
We have introduced a new neural network architecture designed to respect some basic symmetry constraints in particle physics. PELICAN delivers state-of-the-art results in a top-tagging benchmark despite its relatively low complexity. It also shows potential for more complex tasks such as momentum reconstruction, significantly outperforming an established non-ML approach from Ref.~\cite{Kaplan:2008ie}.



\section{Acknowledgements}
We would like to acknowledge Brian Nord and the Deep Skies Lab as a community of multi-domain experts and collaborators who’ve facilitated an environment of open discussion, idea-generation, and collaboration.

\section{Broader Impacts Statement}
This work will potentially have a positive impact on basic physics research: Specifically through the use of PELICAN as a method in particle physics measurements, and perhaps more broadly in providing further inspiration for the use of symmetry-respecting/physically-constrained neural network architectures in scientific research. PELICAN's ability to not only tag particles, but also accurately reconstruct their 4-momenta, opens up possibilities of improving precision measurements of the Standard Model and searches for new physics. This basic research, in turn, will likely continue to yield positive impacts through the innovation and development of new technologies that it drives.
Beyond its potential influence in furthering basic high-energy physics research, this work is unlikely to have other societal impacts (good or bad), as PELICAN does not have clear direct applications outside of physics research, and in fact the use of symmetry-preserving architectures is already present to some extent in industry and government applications -- specifically, the use of convolutional neural networks for image recognition.

\setlength\bibhang{0pt} 
\setlength\bibitemsep{2pt} 
\setstretch{0.9} 
\setquotestyle{english}
\printbibliography[heading=bibintoc]
\setstretch{1} 

\end{document}